\documentclass[twocolumn,aps,floatfix,showpacs,superscriptaddress,pra]{revtex4-1}
\usepackage[T1]{fontenc}
\usepackage{amsmath}
\usepackage{graphicx}
\usepackage[normalem]{ulem}
\usepackage{}

\begin{document}

\title{Effective three-body interactions for bosons in a double-well confinement}

\author{Jacek Dobrzyniecki}
\email{Jacek.Dobrzyniecki@ifpan.edu.pl}
\affiliation{\mbox{Institute of Physics, Polish Academy of Sciences, Aleja Lotnikow 32/46, PL-02668 Warsaw, Poland}}
 
\author{Xikun Li}
\affiliation{Department of Physics and Astronomy, Aarhus University, DK-8000 Aarhus C, Denmark}

\author{Anne E. B. Nielsen}
\altaffiliation{On leave from: Department of Physics and Astronomy, Aarhus University, DK-8000 Aarhus C, Denmark}
\affiliation{Max-Planck-Institut f\"{u}r Physik komplexer Systeme, D-01187 Dresden, Germany}

\author{Tomasz Sowi{\'n}ski}
\affiliation{\mbox{Institute of Physics, Polish Academy of Sciences, Aleja Lotnikow 32/46, PL-02668 Warsaw, Poland}}

\begin{abstract}
When describing the low-energy physics of bosons in a double-well potential with a high barrier between the wells and sufficiently weak atom-atom interactions, one can to a good approximation ignore the high energy states and thereby obtain an effective two-mode model. Here, we show that the regime in which the two-mode model is valid can be extended by adding an on-site three-body interaction term and a three-body interaction-induced tunneling term to the two-mode Hamiltonian. These terms effectively account for virtual transitions to the higher energy states. We determine appropriate strengths of the three-body terms by an optimization of the minimal value of the wave function overlap within a certain time window. Considering different initial states with three or four atoms, we find that the resulting model accurately captures the dynamics of the system for parameters where the two-mode model without the three-body terms is poor. We also investigate the dependence of the strengths of the three-body terms on the barrier height and the atom-atom interaction strength. The optimal three-body interaction strengths depend on the initial state of the system.
\end{abstract}
 
\maketitle
\section{Introduction}

Ultracold atoms in optical lattices, which are used to realize the Bose-Hubbard model, are versatile systems and have received much attention due to the high degree of control of the experimental parameters \cite{bloch2008,zollner2006,zollner2006a,schlagheck2010}. The seminal work by Jaksch \emph{et al.} predicted the superfluid to Mott insulator quantum phase transition \cite{jaksch1998} that has been observed experimentally \cite{greiner2002}. Over a decade now, the scope of the Bose-Hubbard model has been extended to include the effects of excited bands, long-range interactions and interaction-induced tunnelings \cite{Oosten,johnson2009,Larson,Sakmann,dutta2011,Ospelkaus2006,luhmann2012,goral2002,Ni,Ospelkaus,Aikawa,Deiglmayr,SowinskiDutta,zollner2008,zollner2008a,lacki2013,SowinskiPRL,LackiZakrzewski}. Much progress in these \emph{non-standard Hubbard models} has been made in recent years, both in theoretical and experimental studies \cite{dutta2015}. Among the effects mentioned above, the effective three-body interactions are of particular interest \cite{tiesinga2011,valencia2011,sowinski2012,Valencia,cruz2014,avila2014,avila2014a,hincapief2016}. In contrast to the inelastic three-body processes, these effective, coherent three-body interactions are generated by the bosonic two-body interaction-induced virtual excitations to higher bands. Effective three-body interactions explain the rapid damping of revivals observed in experiments which cannot be explained in terms of tunneling or atom losses \cite{johnson2009, greiner2002a, will2010}.

As an ideal model to study these corrections to the standard Hubbard model, the double-well potential is one of the conceptually simplest but important models for describing the Josephson effect in superconducting qubits \cite{cooper2004, simmonds2004,tonel2005,levy2007}, nonlinear self-trapping of Bose-Einstein condensates \cite{shin2004,albiez2005,anker2005,javanainen1986,leblanc2011,gillet2014}, and fermionic mixtures \cite{murmann2015,valtolina2015,zou2014,sowinski2016,tylutki2017}. The stability \cite{winkler2006} and dynamics \cite{folling2007} of repulsive atom pairs with weak interactions in the double-well system have been observed in experiments. For weak interactions, the dynamics of the double-well system can be described by a simplified two-mode model \cite{milburn1997, salgueiro2007,dobrzyniecki2016}. The validity of this approximation of two-mode model relies on the assumption that the on-site interaction is much smaller than the band gap. Therefore, the break-down of this model for the strong interactions is expected as the contributions from the higher bands are not considered \cite{zollner2008,zollner2008a,lacki2013,dobrzyniecki2016}. Inspired by this fact, the three- and higher-body effects should be included as corrections to the two-mode model when the interaction strength is not sufficiently weak.

In this paper, we propose an \emph{extended}, \emph{effective} two-mode Hamiltonian with the aid of two types of effective three-body interactions, and show that this extended two-mode Hamiltonian is sufficient to recover the exact dynamics of three bosons with nearly perfect accuracy. To demonstrate the improvements of the extended two-mode model, we compare the dynamic properties of the exact Hamiltonian with those of the two-mode model and the extended two-mode model. Furthermore, we discuss the behaviors of the optimal values of the two three-body parameters and the dependency on the initial states.

The paper is organized as follows. In Section \ref{sec2}, we introduce the system with a one-dimensional double-well potential and the exact many-body Hamiltonian. In Section \ref{sec3}, we derive the two-mode Hamiltonian and introduce the extended two-mode Hamiltonian. In Section \ref{sec4}, we compare the dynamics of the system predicted by the exact Hamiltonian with that predicted by the two-mode and extended two-mode Hamiltonians. We find that there is a significant improvement of the predictions of the two-mode approximation by incorporating three-body corrections. Subsequently, we investigate the behavior of the optimal values of the three-body parameters. In Section \ref{sec5}, we demonstrate how the optimal value of the three-body parameters depend on the initial state. Finally, we conclude the paper in Section \ref{sec6}.

\section{The model}\label{sec2}
We consider a system of three indistinguishable bosons of mass $m$, confined in a one-dimensional double-well potential $V(x)$. We assume that the particles interact via zero-range forces $V(x-x') = g\delta(x-x')$, where $g$ is the interaction strength. The value of $g$ is directly related to the $s$-wave scattering length and can be controlled experimentally. The Hamiltonian of the many-body system in the second quantization form is given by
\begin{equation}
\label{eq:mb_hamiltonian}
 \hat{\mathcal{H}} = \int \mathrm{d}x \hat{\Psi}^\dagger(x) H_0 \hat{\Psi}(x) + \frac{g}{2} \int \mathrm{d}x \hat{\Psi}^\dagger(x) \hat{\Psi}^\dagger(x) \hat{\Psi}(x) \hat{\Psi}(x),
\end{equation}
where the field operator $\hat{\Psi}(x)$ annihilates a particle at position $x$ and fulfills the bosonic commutation relations $[\hat{\Psi}(x),\hat{\Psi}^\dagger(x')] = \delta(x-x')$, $[\hat{\Psi}(x),\hat{\Psi}(x')] = 0$. The single-particle part of the Hamiltonian has the form
\begin{equation}
\label{eq:1p_hamiltonian}
 H_0 = -\frac{\hbar^2}{2m} \frac{\mathrm{d}^2}{\mathrm{d}x^2} + V(x).
\end{equation}
The potential $V(x)$ is assumed as a superlattice composed by two optical lattices and it has a form
\begin{equation}
\label{eq:potential}
 V(x) =
   V_0\left[ \sin{(kx)}^2 + \lambda \cos{(2kx)}^2 \right],
\end{equation}
where $V_0$ is the lattice depth and $k$ the wave vector of the laser field. The dimensionless parameter $\lambda$ controls the height of the internal barrier in a well forming locally double-well confinement. In further discussion, we express all quantities in natural units of the problem, {\it i.e.}, energies are measured in units of the recoil energy $E_R=\hbar^2 k^2 / 2m$, lengths in units of $k^{-1}$, etc. In the following, we assume that the lattice is very deep ($V_0=15E_R$). Therefore we limit the problem to a single double-well confinement ($|x| \le \frac{\pi}{2}$). (For simplicity, we assume the potential outside the boundaries is infinitely large $V(|x| \geq \frac{\pi}{2})=\infty $.) In consequence, the single-particle eigenproblem $H_0 \Phi_i(x) = \epsilon_i \Phi_i(x)$ can be easily solved numerically, yielding a spectrum of eigenfunctions $\Phi_i(x)$ and their corresponding energies $\epsilon_i$. Due to the symmetry of the problem, all the functions $\Phi_i(x)$ are symmetric or antisymmetric under flipping $x \rightarrow -x$.

For double-well problems, it is convenient to use an alternative single-particle basis of functions localized in a given (left or right) well:
\label{eq:lr_basis}
\begin{subequations}
 \begin{align}
      \varphi_{Li}(x) &= \frac{1}{\sqrt{2}}[\Phi_{2i}(x) - \Phi_{2i+1}(x)],\\
      \varphi_{Ri}(x) &= \frac{1}{\sqrt{2}}[\Phi_{2i}(x) + \Phi_{2i+1}(x)].
  \end{align}
\end{subequations}
In this basis, the single-particle Hamiltonian $H_0$ becomes a block-diagonal matrix with elements given by
\begin{equation}
\label{eq:1p_hamiltonian_lr}
\int\limits_{-\infty}^{\infty} \varphi_{\sigma i}^*(x) H_0 \varphi_{\sigma' j}(x) \mathrm{d}x = \delta_{ij} [ E_i \delta_{\sigma\sigma'} - J_i (1-\delta_{\sigma\sigma'}) ],
\end{equation}
where $\sigma=\{L,R\}$, and
\begin{equation}
 \label{eq:matrixel}
 E_i = \frac{\epsilon_{2i+1}+\epsilon_{2i}}{2}, \quad
 J_i = \frac{\epsilon_{2i+1}-\epsilon_{2i}}{2}.
\end{equation}
By decomposing the field operator $\hat{\Psi}(x)$ in this basis
\begin{equation}
\label{eq:decomposition}
 \hat{\Psi}(x) = \sum\limits_i { [\varphi_{Li}(x) \hat{a}_{Li} + \varphi_{Ri}(x) \hat{a}_{Ri}] },
\end{equation}
the many-body Hamiltonian $\hat{\cal H}$ may be written in a simple Bose-Hubbard-like form
\begin{align}
 \label{eq:mb_hamiltonian_lr}
 \hat{\mathcal{H}}= &\sum\limits_i \left[ E_i ( \hat{n}_{Li} + \hat{n}_{Ri} ) - J_i ( \hat{a}^\dagger_{Li} \hat{a}_{Ri} + \hat{a}^\dagger_{Ri} \hat{a}_{Li} ) \right] \nonumber \\
 &+\frac{1}{2} \sum\limits_{ABCD} U_{ABCD} \hat{a}^\dagger_A \hat{a}^\dagger_B \hat{a}_C \hat{a}_D,
\end{align}
where the operator $\hat{a}_{\sigma i}$ annihilates a boson in the state $\varphi_{\sigma i}(x)$ and fulfills bosonic commutation relations $[\hat{a}_{\sigma i},\hat{a}^\dagger_{\sigma' j}] = \delta_{\sigma\sigma'} \delta_{ij}$ and $[\hat{a}_{\sigma i},\hat{a}_{\sigma' j}] = 0$. The number operator is $\hat{n}_{\sigma i} = \hat{a}^\dagger_{\sigma i} \hat{a}_{\sigma i}$, and the indices $A,B,C,D$ represent super-indices $(\sigma,i)$ numbering the single-particle states $\varphi_{\sigma i}(x)$. The interaction amplitudes $U_{ABCD}$ are given by
\begin{equation}
 \label{eq:interactionterm}
 U_{ABCD} = g \int\limits_{-\infty}^\infty \varphi^*_A(x) \varphi^*_B(x) \varphi_C(x) \varphi_D(x) \mathrm{d}x.
\end{equation}

In numerical calculations, the summation over the single-particle basis in the decomposition (\ref{eq:decomposition}) must be limited to some cutoff value $i_{\max}$. Depending on the interaction strength, the cutoff needed for appropriate description of the system varies. We checked that for all cases studied here, the cutoff $i_{\max} = 15$ is sufficient to describe the system exactly since further increase of $i_{\max}$ does not affect the final results significantly.

\section{Effective Hamiltonian approach}\label{sec3}

The Hamiltonian (\ref{eq:mb_hamiltonian_lr}) can be simplified by assuming that the dynamics is limited to the lowest single-particle basis $\varphi_{L0}(x)$ and $\varphi_{R0}(x)$. Under this assumption one neglects all states with $i > 0$ and the decomposition (\ref{eq:decomposition}) becomes simplified to $\hat{\Psi}(x) = \varphi_{L0}(x) \hat{a}_{L0} + \varphi_{R0}(x) \hat{a}_{R0}$. In this two-mode approximation the resulting many-body Hamiltonian reads:
\begin{align}
\label{eq:mb_2m_hamiltonian}
 \hat{\mathcal{H}}_\mathtt{2mode} &= E_0 (\hat{n}_L + \hat{n}_R) - J_0 (\hat{a}^\dagger_L \hat{a}_R + \hat{a}^\dagger_R \hat{a}_L) \\
 &+ \frac{U}{2} \left[\hat{n}_L(\hat{n}_L - 1) + \hat{n}_R(\hat{n}_R - 1)\right] \nonumber \\
 &+ T \left[ \hat{a}^\dagger_L (\hat{n}_L+\hat{n}_R) \hat{a}_R + \hat{a}^\dagger_R (\hat{n}_L+\hat{n}_R) \hat{a}_L \right] \nonumber \\
 &+ \frac{V}{4} \left[(\hat{a}^\dagger_L)^2 (\hat{a}_R)^2 + (\hat{a}^\dagger_R)^2(\hat{a}_L)^2+ 4 \hat{n}_L \hat{n}_R \right], \nonumber
\end{align}
where for convenience we omit index $i$. The interaction terms are denoted by $U = U_{LLLL}$, $T = U_{LLLR}$, $V = 2U_{LLRR}$.

The two-mode approximation (\ref{eq:mb_2m_hamiltonian}) is justified in the case where the level spacing between the ground ($i=0$) and the first excited band ($i=1$) is much greater than the two-body interaction energy per particle. However, as the interaction strength grows, the excited single-particle states start to influence the properties of the system and the two-mode approximation becomes increasingly inaccurate. In consequence, higher-band states have to be taken into account in the decomposition (\ref{eq:decomposition}). 

A direct addition of higher-band states to the model inevitably increases the complexity of numerical calculations. To overcome this difficulty one can tread different ways to include their effects while staying within a two-mode framework. For example one can effectively change the shape of single-particle orbitals and tailor them to the specific initial state and interactions \cite{DobrzynieckiPLA}. Alternatively, one can leave the lowest orbitals unchanged but include appropriate corrections induced by interactions to the many-body Hamiltonian \eqref{eq:mb_2m_hamiltonian}. Johnson {\it et al.} showed that appropriate corrections can be obtained straightforwardly via perturbation theory \cite{johnson2009,johnson2012}. The resulting correction terms include not only modifications to the two-body interaction strength, but also the appearance of effective $N$-body interactions. These corrections can be explained as arising from density-dependent modifications to the on-site orbitals \cite{smerzi2003,bissbort2012,li2006}, which cause the individual terms in the Hamiltonian to become dependent on the particle number \cite{dutta2011}. 
Typically, for static situations it is sufficient to take into account only corrections to the local interaction energy \cite{pielawa2011,sowinski2012,valencia2011,cruz2014,avila2014,hincapief2016,avila2014a}. However, when the initial state of the system is far from the ground state, other interaction processes are sensitive to interactions with higher bands and their amplitudes have to be treated as occupation-dependent \cite{bissbort2012}.

Inspired by these different approaches, here we propose an intermediate approach which is very efficient when the problem of $N=3$ and more bosons in a double-well potential is considered. Instead of calculating all necessary  corrections to all two-body interaction terms in \eqref{eq:mb_2m_hamiltonian} and taking into consideration all the effective three-body interactions, we improve the 'bare' two-mode description \eqref{eq:mb_2m_hamiltonian} by adding only two appropriately chosen three-body terms which encompass all the relevant processes induced by higher bands. These terms can be interpreted as: (\emph{i}) the on-site three-body interaction which effectively introduces corrections to the energy caused by perturbative change of the orbital's density profile; (\emph{ii}) the three-body interaction-induced tunneling which takes into account an effective modification of tunnelings due to tunnelings in higher occupied orbitals. The extended two-mode Hamiltonian reads:
\begin{align}
\label{eq:3body_hamiltonian}
 \hat{\mathcal{H}}_\mathtt{eff} &= \hat{\mathcal{H}}_\mathtt{2mode}  \\
 &+ \frac{W}{6} \left[ \hat{n}_L(\hat{n}_L - 1)(\hat{n}_L - 2) + \hat{n}_R(\hat{n}_R - 1)(\hat{n}_R - 2) \right] \nonumber \\
 &+ \frac{Q}{2} \left[(\hat{a}^\dagger_L \hat{a}^\dagger_L \hat{a}^\dagger_L \hat{a}_L \hat{a}_L \hat{a}_R + \hat{a}^\dagger_R \hat{a}^\dagger_R \hat{a}^\dagger_R \hat{a}_R \hat{a}_R \hat{a}_L) + h.c.\right] \nonumber
\end{align}

In the following we consider different interactions $g$ and double-well confinements $\lambda$ for which the effective Hamiltonian \eqref{eq:3body_hamiltonian} can be used to describe the dynamics of the system correctly.

\section{The dynamics} \label{sec4}
\label{sec:results}
It is quite natural that the magnitudes of the effective three-body parameters $W$ and $Q$ depend on the potential barrier height $\lambda$ as well as the two-body interaction strength $g$. They may also depend on the initial state $|\mathtt{ini}\rangle$ in which the system is prepared. For a given initial state $|\mathtt{ini}\rangle$, the exact time evolution of the system state $|\mathbf{\Psi}(t)\rangle$ is given by $|\mathbf{\Psi}(t)\rangle = \exp{(-i\hat{\mathcal{H}}t)} |\mathtt{ini}\rangle$. This state can be directly compared with the dynamics carried by the effective Hamiltonian $|{\psi}(t)\rangle = \exp{(-i\hat{\mathcal{H}}_{\mathtt{eff}}t)} |\mathtt{ini}\rangle$. In this way one can define the time-dependent fidelity $\mathcal{F}(t)$ as the overlap between temporal states of the system
\begin{equation}
\label{eq:fidelity}
\mathcal{F}(t) = \lVert \langle \mathbf{\Psi}(t) | \psi(t) \rangle \rVert^2.
\end{equation}
Of course the fidelity \eqref{eq:fidelity} varies in time. Therefore, instead of considering the fidelity ${\cal F}(t)$, we focus on some initial time period $0\leq t\leq\tau$ and we find the minimal value ${\cal F}_\mathrm{min}=\mathrm{min}\{{\cal F}(t): 0\leq t\leq\tau\}$ reached by the fidelity in this period. In the following we choose $\tau=12\pi\hbar/J_0$, {\it i.e.}, a quite large multiple of the natural time scale $\pi \hbar/J_0$ related to the time which is needed for a non-interacting system to return to its original state.

\begin{figure}
\includegraphics[width=1\linewidth]{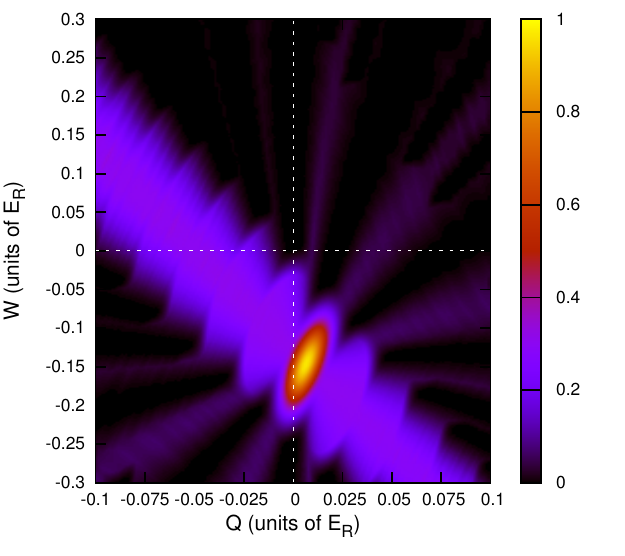}
\caption{The fidelity $\mathcal{F}_{\mathrm{min}}$ as a function of the coefficients $W$ and $Q$, for the case $\lambda = 1$, $g = 0.7 E_R/k$. A single, clear peak is visible, with the maximum fidelity corresponding to the point $W_0 \approx -0.0147, Q_0 \approx 0.0068 $.}
 \label{fig:fidelityscan}
\end{figure}

We choose $Q$ and $W$ to maximize ${\cal F}_\mathrm{min}$, thus the effective two-mode model recovers the exact dynamics of the system as much as possible. The simplest way to find the optimal values of $W$ and $Q$, which we denote $W_0$ and $Q_0$, for a system with given parameters $\lambda$ and $g$ is to examine properties of ${\cal F}_\mathrm{min}$ as a function of $W$ and $Q$. From our numerical calculations it follows that ${\cal F}_\mathrm{min}(W,Q)$ has a global maximum for a well defined pair $(W_0,Q_0)$. For example, in Fig.\ \ref{fig:fidelityscan} we plot the minimal fidelity ${\cal F}_\mathrm{min}$ for a single choice of the parameters ($\lambda=1$ and $g=0.7$) and the initial state $|\mathtt{0,3}\rangle = \frac{1}{\sqrt{3!}}(\hat{a}^\dagger_{R0})^3 |\mathtt{vac}\rangle $, {\it i.e.}, the state with all three bosons initially localized in the right well. It should be stressed that the values of $W_0$ and $Q_0$ are not necessarily directly related to the actual three-body corrections derivable in perturbation theory. Rather, these values, when optimized, effectively encompass several different $N$-body corrections.

\begin{figure}
 \includegraphics[width=1\linewidth]{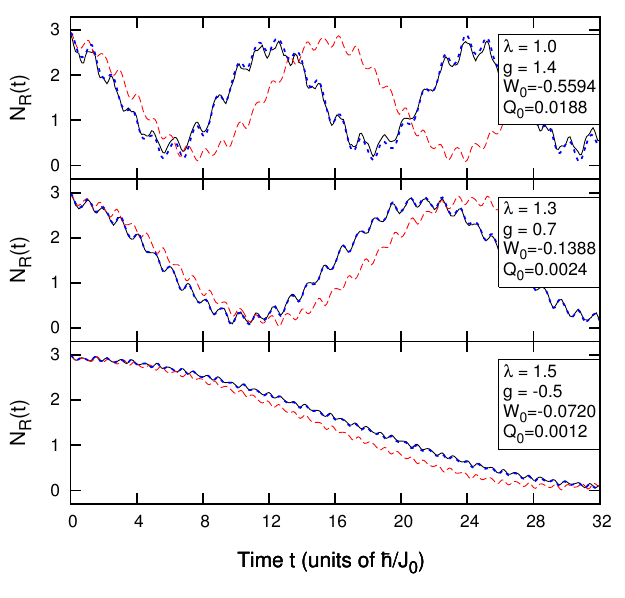}
 \caption{The time evolution of the population in the right well $N_R(t)$ as governed by the full Hamiltonian $\hat{\mathcal{H}}$ (black, solid line), the standard two-mode Hamiltonian $\hat{\mathcal{H}}_\mathtt{2mode}$ (red, dashed), and the extended two-mode Hamiltonian with three-body interactions $\hat{\mathcal{H}}_\mathtt{eff}$ (blue, dotted). The optimal three-body parameters $W_0$, $Q_0$ are determined by maximising ${\cal F}_\mathrm{min}$. Note, that appropriate three-body terms lead to very good predictions of the exact dynamics. Coefficients $(W_0,Q_0)$ and interaction strength $g$ are given in units of $E_R$ and $E_R/k$, respectively.}
 \label{fig:population-optical}
\end{figure}

To show that incorporating three-body corrections indeed significantly increases accuracy of the two-mode approximation, in Fig.\ \ref{fig:population-optical} we compare the population of the right well $N_R(t)$ predicted by the exact Hamiltonian $\hat{\cal H}$
\begin{equation}
\label{eq:rightpopulation}
N_R(t) = \int\limits_0^{\infty} \langle \mathbf{\Psi}(t) | \hat{\Psi}^\dagger(x) \hat{\Psi}(x) | \mathbf{\Psi}(t) \rangle \mathrm{d}x
\end{equation}
with corresponding quantities predicted by $\hat{\cal H}_\mathtt{2mode}$ and $\hat{\cal H}_\mathtt{eff}$.

In the considered range of interactions and barrier heights, the population $N_R(t)$ displays a characteristic oscillatory behavior. When the interactions are not negligible, the dynamics of the traditional two-mode model (red, dashed line) clearly deviates from the exact dynamics (black line), underestimating (for repulsive interactions) or overestimating (for attractive interactions) the oscillation frequency. However, when the three-body corrections are incorporated into the two-mode model, there is a significant improvement of the predictions. The extended two-mode model with three-body corrections (blue line) recovers the exact time evolution of the population with nearly perfect accuracy, for both attractive and repulsive interactions.

\begin{figure}
 \includegraphics[width=1\linewidth]{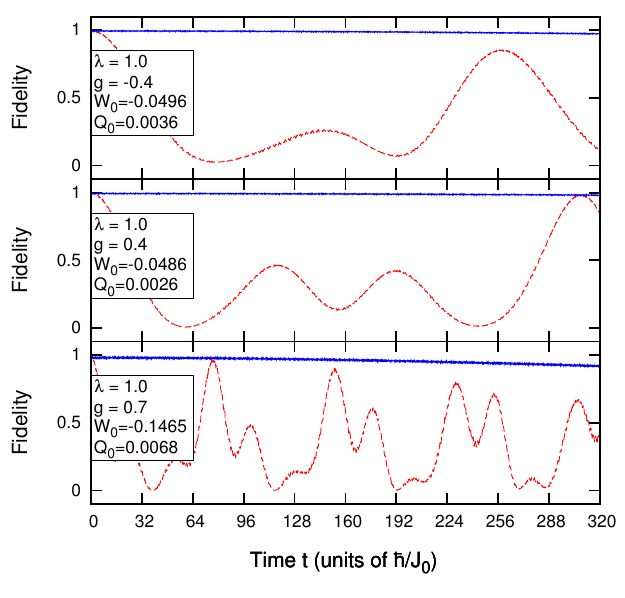}
 \caption{ The time evolution of the fidelity $\mathcal{F}(t)$ for the standard two-mode Hamiltonian $\hat{\mathcal{H}}_\mathtt{2mode}$ (red, dashed) and the extended two-mode Hamiltonian with three-body interactions $\hat{\mathcal{H}}_\mathtt{eff}$ (blue, solid), for various interaction strengths in a shallow well ($\lambda = 1.0$). The fidelity in the extended $\hat{\mathcal{H}}_\mathtt{eff}$ model stays near $1$ even for long times, unlike the fidelity of the standard $\hat{\mathcal{H}}_\mathtt{2mode}$ model. Coefficients $(W_0,Q_0)$ and interaction strength $g$ are given in units of $E_R$ and $E_R/k$, respectively.}
 \label{fig:fidelity-optical}
\end{figure}

As an additional demonstration of the improvement granted by the extended model, in Fig.\ \ref{fig:fidelity-optical} we show the evolution of $\mathcal{F}(t)$ over a long timescale, for various interaction strengths with fixed lattice depth $\lambda = 1.0$. In the traditional two-mode model, the overlap of the approximate system state with the exact state drops to zero fairly quickly. While $\mathcal{F}(t)$ subsequently goes through revivals, the average fidelity over a long time is significantly below $1$. The extended model with three-body corrections, on the other hand, recovers the exact system state very closely even for long times, and the fidelity drops very slowly.

\begin{figure}
 \includegraphics[width=1\linewidth]{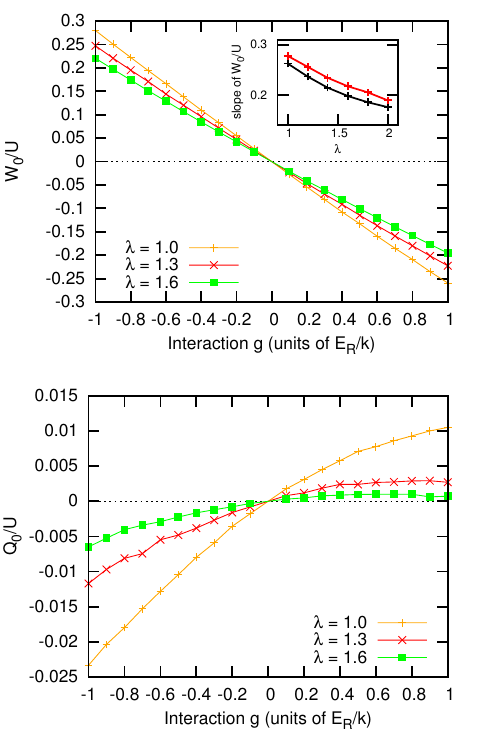}
 \caption{The values of $W_0/U$ (top) and $Q_0/U$ (bottom) as a function of interaction $g$ and barrier height $\lambda$ for a system initially prepared in the state $|\mathtt{0,3}\rangle$. In the considered range of interactions a linear regression of $W_0/U$ indicates that $W_0$ is quadratic in $g$. Inset: the slope of a linear fit to $W_0/U$ for different barrier heights $\lambda$, for repulsive (black lower line) and attractive (red upper line) interactions. Note that the slope decreases with increasing $\lambda$. }
 \label{fig:wq-linear}
\end{figure}

The optimal values $W_0$ and $Q_0$ obtained for different interactions $g$ and barriers $\lambda$ are shown in Fig.\ \ref{fig:wq-linear}. For convenience, they are expressed in units of the on-site two-body interaction energy $U$ as a natural point of comparison of interactions in a given system. As shown in Fig.\ \ref{fig:wq-linear}, $W_0/U$ scales linearly with the interaction strength $g$, thus $W_0$ is an almost precisely quadratic function of $g$ ($U$ is proportional to $g$). The behavior of $W_0/U$ is consistent with theoretically expected trends in $W_0$, as the dominant contribution to the on-site three-body correction originates from second-order virtual processes \cite{johnson2009}. Note, however, that $W_0/U$ has different slopes in the repulsive and the attractive regimes. This fact may be viewed as a result of creation of bounding pairs in the attractive regime leading directly to nonnegligible corrections from higher order processes.

We notice that the on-site three-body interaction $W_0$ is negative in the repulsive interaction ($g>0$) as well as in the attractive interaction case ($g<0$). This result can be explained intuitively when one considers how the $N$-body corrections change the shape of on-site orbitals. Repulsive interactions lead effectively to a broadening of the single-particle orbital. In consequence the value of the on-site interaction term $U$, assuming that the shapes of the orbitals are interaction-independent, overestimates the interaction energy. Therefore, the on-site three-body correction has to be negative. In contrast, for attractive interactions the two-body term $U$ underestimates the energy (understood as its absolute value) since in this case orbitals become squeezed. As a result, the on-site three-body correction $W$ has to be also negative.

For completeness, we also plot the slope of $W_0/U$, as a function of $\lambda$ in the vicinity of $g=0$, {\it i.e.}, $|g| \in [0,1]$ (inset in Fig.\ \ref{fig:wq-linear}). The slope decreases with rising barrier height, indicating that the magnitude of the on-site three-body correction (when compared to the two-body terms) decreases as the barrier becomes higher. This observation is consistent with the fact that an influence of higher bands is smaller for deeper lattices since the band-gap between the ground and excited states becomes larger \cite{dobrzyniecki2016}.

The behavior of the tunneling induced by three-body interactions of $Q_0/U$ is quite different. As it is seen in the bottom panel in Fig.\ \ref{fig:wq-linear}, its magnitude grows as the interaction strength increases and, in contrast to $W_0$, it is quite far from the quadratic behaviour. Within the examined range of experimental regimes the value of $Q_0/U$ is always  positive. It means that a strength of an effective single-particle tunneling decreases (single-particle tunneling has opposite sign). This result is fully consistent with findings in \cite{luhmann2012}, where, for bosons in an optical lattice, the perturbative correction was found to decrease the tunneling under similar conditions.

The magnitude of the tunneling correction $Q_0$ is in general much smaller than the on-site three-body correction $W_0$. For a low potential barrier ($\lambda = 1$) $Q_0$ is one order of magnitude smaller than $W_0$. For higher barriers the ratio quickly drops, and for $\lambda = 1.6$ $Q_0$ is almost two orders of magnitude smaller than $W_0$. This trend is also consistent with intuitive expectations, since the role of first-order tunneling effects should naturally decrease as the height of the inter-well barrier grows \cite{folling2007}.

 \begin{figure}
 \includegraphics[width=1\linewidth]{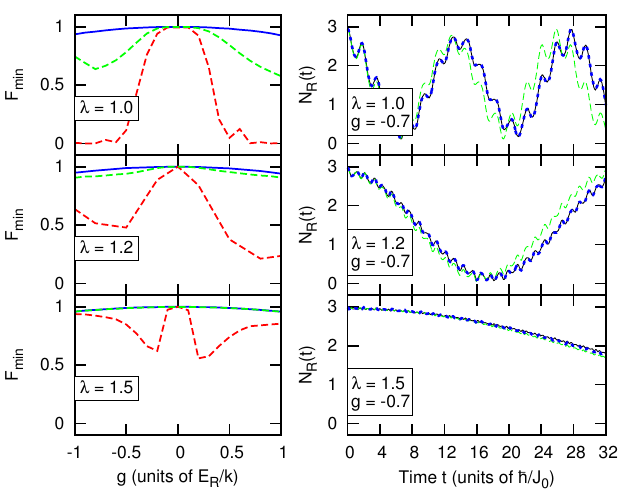}
 \caption{Left panel: The fidelity $\mathcal{F}_{\mathrm{min}}$ as a function of $g$ when the dynamics is governed by the standard two-mode Hamiltonian $\hat{\mathcal{H}}_\mathtt{2mode}$ (red dashed (lower) line), the extended two-mode Hamiltonian $\hat{\mathcal{H}}_\mathtt{eff}$ (blue solid line), and the extended two-mode Hamiltonian $\hat{\mathcal{H}}_\mathtt{eff}$ without interaction-induced tunneling ($Q=0$) (green dashed (upper) line). For $Q=0$, a value of $W_0$ was chosen that maximizes $\mathcal{F}_{\mathrm{min}}$ within the $Q=0$ constraint. Results are shown for various barrier heights $\lambda$. For shallow wells the fidelity of the standard two-mode model $\hat{\mathcal{H}}_\mathtt{2mode}$ drops rapidly close to 0. Three-body on-site interaction improves significantly the accuracy of the model. For deep wells the tunneling correction controlled by $Q_0$ does not play a significant role. Right panel: Time evolution of the right-well population $N_R(t)$ as governed by the full Hamiltonian $\hat{\mathcal{H}}$ (black solid line), the Hamiltonian $\hat{\mathcal{H}}_\mathtt{eff}$ with (blue dotted line) or without (green dashed line) the single-particle tunneling correction term. The inaccuracy that results from omitting the tunneling correction is apparent for shallow wells.}
 \label{fig:fidelity-Q}
\end{figure}

Although $Q_0$ is one (or two) order(s) of magnitude smaller than $W_0$, in general it cannot be neglected when the dynamical properties of the system are studied. This fact is demonstrated in Fig.\ \ref{fig:fidelity-Q}. In the left panel we compare the fidelity $\mathcal{F}_{\mathrm{min}}$ for various approximations of the Hamiltonian $\hat{\cal H}$ as a function of interaction strength $g$. The red, dashed line shows the fidelity $\mathcal{F}_{\mathrm{min}}$ obtained for the basic two-mode model $\hat{\mathcal{H}}_\mathtt{2mode}$. The blue line shows $\mathcal{F}_{\mathrm{min}}$ when both three-body corrections are incorporated and the system is described by the Hamiltonian $\hat{\mathcal{H}}_\mathtt{eff}$. The green line shows $\mathcal{F}_{\mathrm{min}}$ for a specific effective Hamiltonian $\hat{\mathcal{H}}_\mathtt{eff}$ assuming that the tunneling induced by three-body interactions can be neglected, {\it i.e.}, $Q\equiv 0$ and the value of $W_0$ is chosen anew to optimize $\mathcal{F}_{\mathrm{min}}$ under this condition. As it is seen, the three-body correction controlled by $Q_0$ becomes important for shallow barriers.

To give a better understanding of this observation, in the right panel in Fig.\ \ref{fig:fidelity-Q} we qualitatively compare $N_R(t)$ predicted by different approximations for three different lattice depths. As it is seen, for a small barrier height ($\lambda=1.0$), there is an increasing discrepancy between the models. Indeed, the difference becomes much smaller when the barrier height is increased (see caption of Fig.\ \ref{fig:fidelity-Q} for details).

\section{Role of the initial state } \label{sec5}

\begin{figure}
 \includegraphics[width=1\linewidth]{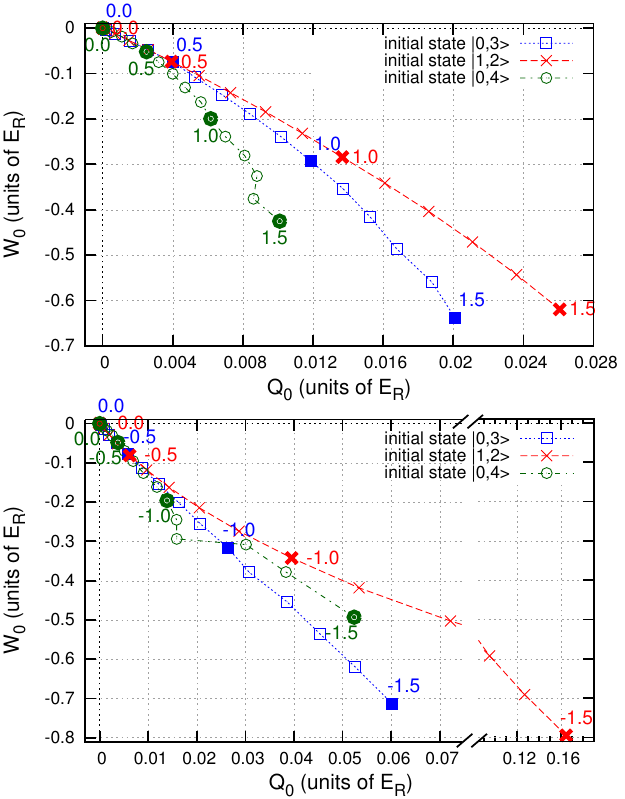}
 \caption{The values of $W_0$ and $Q_0$ obtained for different initial states $|\mathtt{0,3}\rangle$ (blue line, squares), $|\mathtt{1,2}\rangle$ (red line, crosses), and $|\mathtt{0,4}\rangle$ (green line, circles) and different values of repulsive (top panel) and attractive (bottom panel) interactions ($g$ is in the range $-1.5$ to $1.5$ with a step size of 0.1, in units of $E_R/k$). All results are obtained for a specific height of the barrier, $\lambda = 1$.}
 \label{fig:wq-differentstates}
\end{figure}

Up to now, all results were obtained for a specific initial state of three bosons $|\mathtt{0,3}\rangle$, {\it i.e.}, the state with three bosons occupying a single well. If we interpret the three-body corrections $W_0$ and $Q_0$ as the actual magnitudes of effective three-body interactions derived from underlying physical processes, they should be almost insensitive to the initial state. However, as noted previously, in our approach these terms encompass the effects of several different corrections and they do not correspond exactly to specific phenomena. It is therefore possible that their optimal values depend on the initial state. A natural question arises regarding the robustness of our effective three-body correction approach when different initial states and higher number of particles are considered.

Therefore, in the following, we examine the dynamical properties for two other initial states: $|\mathtt{1,2}\rangle = \frac{1}{\sqrt{2!}} (\hat{a}^\dagger_{R0})^2 \hat{a}^\dagger_{L0} |\mathtt{vac}\rangle$ and $|\mathtt{0,4}\rangle = \frac{1}{\sqrt{4!}}(\hat{a}^\dagger_{R0})^4 |\mathtt{vac}\rangle$. In both cases we optimise the three-body corrections $W$ and $Q$ and compare them to those obtained previously for the state $|\mathtt{0,3}\rangle$. We find that, while for other initial states it is still possible to attain a significant improvement in the fidelity $\mathcal{F}_{\mathrm{min}}$, the optimal values $W_0$ and $Q_0$ are in general different depending on the initial state.

The results for the double-well potential with barrier $\lambda = 1$ are summarised in Fig.\ \ref{fig:wq-differentstates}. The plotted data points in $(W_0,Q_0)$ space correspond to various values of $g$ (top and bottom panel are for repulsive and attractive interactions, respectively). The three different colors correspond to the different initial states of the system (blue for $|\mathtt{0,3}\rangle$, red for $|\mathtt{1,2}\rangle$, and green for $|\mathtt{0,4}\rangle$ ). First, we compare the results for the two initial states $|\mathtt{0,3}\rangle$ and $|\mathtt{1,2}\rangle$ with the same total number of particles. For weak interactions $|g|$, the resulting three-body corrections are similar for both states, regardless of the initial state. Note that for stronger interactions the difference between the on-site three-body correction $W_0$ for the two initial states remains almost negligible (for $g=1.5$ the difference is of the order of $10\%$). In contrast, the tunnelling correction $Q_0$ is strongly dependent on the initial state. There is only a finite range of weak interactions within which the two values of $Q_0$ are still reasonably similar. For the case considered, \emph{i.e.}, barrier height $\lambda = 1$, this range is approximately $-0.6 \le g \le 1.0$. Beyond this range, the values of $Q_0$ are essentially different for different initial states, and in the most extreme case studied, $g = -1.5$, the difference is of the order of $300\%$.

The situation is even more complicated in the case of four bosons, $|\mathtt{0,4}\rangle$ (green lines). Indeed, for $N=4$ the optimal value $W_0$ as well as $Q_0$ is essentially different from those obtained in the case of three bosons. This means that, although three-body corrections $W_0$ and $Q_0$ can be nicely fitted to obtain appropriate predictions for the dynamics in different scenarios, their values are not universal and depend on the initial state as well as on the interaction strength. In general, the universal pair $(W_0,Q_0)$ leading to high fidelity of the dynamics regardless of initial state does not exist. However, within a limited range of interactions $g$, the same pair $(W_0,Q_0)$ can be used equally well for the two initial states $|\mathtt{0,3}\rangle$ and $|\mathtt{1,2}\rangle$. From this point of view, the approximate Hamiltonian $\hat{\cal H}_\mathtt{eff}$ should be treated as an effective state-dependent description, not derivable solely from the fundamental many-body Hamiltonian.

\section{Conclusions} \label{sec6}

We studied the dynamics of a few-boson system confined in a one-dimensional double-well potential. We introduced an extended two-mode model by taking into account two effective three-body interactions, {\it i.e.}, on-site three-body interaction and three-body interaction-induced tunneling. Rather than including exact values of perturbative corrections to the interactions, we encompass different effects of multiple corrections by adding only two effective interaction terms. Their optimal values were obtained by maximizing the minimal value of the fidelity which measures the overlap between the time evolution of the states governed by the exact Hamiltonian and the extended two-mode model. We compared the dynamics of the population in the right well for different Hamiltonians, and found that the exact dynamics, which clearly deviated from the dynamics of the two-mode model, were well approximated by the extended two-mode model with nearly perfect accuracy. The behaviors of the optimal values of the two three-body parameters were investigated. In particular, the quadratic form in on-site three-body interaction agrees with the theoretical prediction. By studying the dynamics governed by the extended two-mode models with and without interaction-induced tunneling (which was small compared to the on-site three-body interaction), we found that the discrepancy between these two models was increasing with stronger interactions, thus the necessity of interaction-induced tunneling was validated. Finally, we examined the robustness of the two three-body parameters to different initial states and observed that both of them are sensitive to the initial states when the interaction strength is strong. 

In order to fully verify the universality of our model, we also tested it for a differently shaped double-well potential. It is given by a combination of a harmonic oscillator well of frequency $\omega$ with a Gaussian-shaped barrier of height regulated by parameter $\lambda$. Assuming harmonic-oscillator units, where energy is given in $\hbar \omega$ and position in $\sqrt{\hbar/m\omega}$, the potential can be written as $V(x) = x^2/2 + \lambda \exp{\left(-x^2/2\right)}$. We have confirmed that for a system in this potential our model still functions properly. It is still possible to find optimal values of $(W,Q)$ such that the Hamiltonian $\hat{\mathcal{H}}_\mathtt{eff}$ recovers the exact dynamics properly, showing that the usefulness of the model is independent of the fine details of the potential shape.

\begin{acknowledgments}
This work has in part been supported by the Villum Foundation. J.D. and T.S. acknowledge support from the (Polish) National Science Center Grant No. 2016/22/E/ST2/00555. X.L. thanks the Max Planck Institute for the Physics of Complex Systems for hospitality during visits to the institute. 
\end{acknowledgments}

\end{document}